\documentclass[preprint,12pt]{elsarticle}

\usepackage{amssymb}
\usepackage[hidelinks]{hyperref}
\usepackage{lineno}
\usepackage{amsmath}
\usepackage{float}
\usepackage{subcaption}
\usepackage{mathtools, nccmath}
\usepackage[bb = px]{mathalfa}
\usepackage{bbm}
\usepackage{amssymb}
\usepackage{gensymb}
\usepackage{graphicx}
\usepackage{dcolumn}
\usepackage{rotating}
\usepackage{color}
\usepackage{times}
\usepackage{url}
\usepackage{epstopdf,multirow}
\usepackage{geometry}
\usepackage{pstricks}
\usepackage{fancyhdr}
\usepackage{booktabs}
\usepackage{grffile}
\usepackage[font={small}]{caption}
\usepackage{subcaption}
\usepackage{multirow}
\usepackage{float}
\usepackage{sectsty}
\usepackage{siunitx}

\usepackage[]{lineno}
\usepackage[below]{placeins}
\usepackage{setspace}
\usepackage{needspace}

\newcommand{\intu}{\medint\int\displaylimits}

\makeatletter
\def\ps@pprintTitle{%
}
\makeatother

\begin{document}
\pagestyle{plain}
\renewcommand{\arraystretch}{1.3}


\onehalfspacing

\begin{frontmatter}		
\title{Moment-Based Transient Network Models for Finite-Strain Viscoelasticity}
\author[llnl]{Stephen T. Castonguay\corref{cor1}}
\author[llnl]{Pratanu Roy}
\author[llnl]{Sylvie Aubry}
\cortext[cor1]{Corresponding author. Email: castonguay1@llnl.gov}
\address[llnl]{Lawrence Livermore National Laboratory, Livermore, CA 94551, USA}

\begin{abstract}
A viscoelastic framework is developed through a moment reduction of the hereditary integral defining the free energy of transient polymer networks. Expanding the chain energy about the mean history invariant produces a hierarchy of tensorial moments representing the network formation history. For energies depending on the relative first invariant, the leading approximation recovers Lubliner's model, providing a physical interpretation of its second-rank history tensor as an average of the inverse deformation metrics at chain formation. The next approximation introduces a fourth-rank tensor that accounts for variability among formation configurations. These moments admit equivalent rate forms and efficient recurrence relations, while their evolution permits the establishment of sufficient conditions for dissipation. Comparisons with an exact hereditary representation of the three-term Arruda–Boyce energy demonstrate that the first-moment approximation captures the overall response, while retaining the second moment improves accuracy, particularly under complex loading histories. We also incorporate non-exponential relaxation through distributed detachment kinetics described by the Kohlrausch–Williams–Watts (KWW) function. Assuming thermorheological simplicity, calibration to thermally accelerated butyl-rubber relaxation measurements yields excellent agreement across wide ranges of time and temperature. The calibrated continuous rate spectrum is then discretized into a finite Prony series using equal-probability bins. The proposed framework provides a practical representation of nonlinear transient network behavior that separates the approximation of deformation history from the discretization of the continuous rate spectrum.
\end{abstract}
%

\end{frontmatter}	
\section{Introduction}
Transient network models describe the relaxation of polymers and the evolution of their natural configuration through the continual breaking and reformation of bonds. As new bonds form, they establish network populations that are stress-free in their configurations at formation and contribute to the subsequent mechanical response. The free energy is therefore expressed as a hereditary integral over surviving network populations, each deformed relative to its configuration at formation. This theory dates back to the pioneering works of Green and Tobolsky \cite{green1946new}, who postulated that stress relaxation and residual strain occur due to crosslink breakage and reformation. Their model assumed an incompressible material, as well as a Gaussian distribution of end-to-end chain vectors. They also showed how this model could be recast in rate form for an internal second-rank history tensor. 

Lubliner \cite{Lubliner1985} subsequently developed a viscoelastic model based on a multiplicative decomposition of the deformation gradient into elastic and inelastic parts. He adopted the linear evolution law of Green and Tobolsky’s internal variable and identified that variable with the inverse inelastic right Cauchy--Green tensor within the multiplicative framework. His model accommodates material compressibility, more general elastic free energies, and multiple relaxation times. However, the assumption of isochoric inelastic deformation requires the internal variable to have unit determinant, which the linear evolution does not preserve. Lubliner explicitly recognized this, and interpreted the linear law as an approximation to some constraint-preserving nonlinear evolution within the multiplicative framework. Reese and Govindjee \cite{ReeseGovindjee1998FiniteViscoelasticity} subsequently developed such a nonlinear formulation and showed it recovered Lubliner's model under appropriate linearization.

More recently, Vernerey et al. \cite{Vernerey2017TransientNetworks} developed a statistical transient-network theory based on an evolving distribution of chain end-to-end vectors. For Gaussian-chain elasticity, its second moment, termed the chain-distribution tensor, obeys a closed evolution equation. Vernerey \cite{Vernerey2018KineticTheory} extended this framework to nonlinear Langevin-chain elasticity using a mean-field approximation. 
Hui et al. \cite{Hui2021DynamicCrosslinks} compared this framework with hereditary integral formulations, identifying them as Eulerian and Lagrangian counterparts and establishing their equivalence under a Gaussian distribution, with the Green–Tobolsky theory providing the classical Lagrangian form. Their comparison highlights the computational costs associated with each, as hereditary integral evaluation requires retaining the network formation history, while the statistical formulation requires resolving a chain-configuration distribution at each material point.

Various forms of transient network theory have been applied to a wide range of materials, including rubbers \cite{budzien2008new,linder2011micromechanically}, gels \cite{long2014time,hui2012constitutive}, vitrimers \cite{karim2025constitutive}, and associating polymer fluids \cite{wang1992transient,tanaka1992viscoelastic}. Their ability to span solid-like and fluid-like behavior through network renewal allows them to describe viscoelasticity, chemical aging, permanent set, and viscous flow within a common framework \cite{tanaka1992viscoelastic,septanika1998application}. Although the underlying physical mechanisms differ, these behaviors can be represented through appropriate choices of network kinetics and elastic free energy.

Previously, we derived exact recurrence relationships for hereditary integral, transient network models with suitable decomposable free energies \cite{castonguay2026hereditary}. This focused on phenomenological macroscopic energy functions, using the transient network interpretation without explicitly connecting the elastic response to chain-level models. The recurrence relationships eliminate the need to store the entire history, but for invariant-based energies this applies primarily to polynomial forms, with higher polynomial orders requiring progressively higher-rank history tensors. Here, we develop a reduced representation by expanding the free energy about the mean history invariant. 
The leading approximation represents the transient network history through a second-rank tensor, while the next correction involves a fourth-rank tensor which accounts for the variance among formation configurations. We illustrate the reduction using the widely used Arruda–Boyce eight-chain model \cite{arruda1993three}, which connects macroscopic strain stiffening to finite chain extensibility. For first-invariant-based energies like Arruda-Boyce, the leading term recovers the corresponding Lubliner form. This gives Lubliner's history variable a direct alternative interpretation as a
moment of the network formation history.

Non-exponential relaxation is incorporated through distributions of first-order detachment processes. We consider the Kohlrausch–Williams–Watts (KWW) relaxation function, which admits an equivalent representation as a continuous spectrum of exponential processes \cite{pollard1946representation,johnston2006stretched}. In transient-network theory, these processes correspond naturally to chain populations with different detachment rates. Temperature dependence is introduced through a common Vogel–Fulcher–Tammann (VFT) relationship for the detachment rates, which recovers Arrhenius kinetics as a special case and yields the Williams–Landel–Ferry (WLF) form for time–temperature superposition.



The continuous KWW rate spectrum is discretized into a finite Prony series for efficient numerical implementation. The proposed framework is then assessed through two sets of examples. First, comparisons with the exact hereditary response under uniaxial, cyclic, and non-proportional loading quantify the accuracy of the first- and second-moment approximations. Second, the temperature-dependent KWW response is calibrated to thermally accelerated rubber relaxation measurements, followed by an assessment of the accuracy of its Prony discretization. Together, these examples quantify the separate errors introduced by approximating deformation history through tensorial moments and distributed kinetics through a finite set of relaxation rates.

\section{Transient network theory and moment reduction} \label{sec:TNTheory}
For simplicity, we take the volumetric response to be hyperelastic, while the isochoric response is governed by a transient network. The specific free energy is therefore decomposed as
\begin{equation}
	\Psi = \Psi_{\mathrm{vol}} + \Psi_{\mathrm{iso}} .
\end{equation}
A network population formed at time $s$ contributes to the energy at time $t$ through the chains that remain active and their deformation relative to the configuration at formation. Let $\xi(t,s)$ denote the number of active chains per unit mass and per unit formation time, and $\psi_c$ the free energy per chain. The deformation gradient $\mathbf F(t)$ is measured relative to the original reference configuration, with $\mathbf C(t)=\mathbf F^{\mathsf T}(t)\mathbf F(t)$ and $J(t)=\det\mathbf F(t)$. Since each network is stress-free in its formation configuration, its subsequent deformation is measured by $\mathbf F(t,s)=\mathbf F(t)\mathbf F^{-1}(s)$, with relative Jacobian $J(t,s)=J(t)/J(s)$. The corresponding relative right Cauchy–Green tensor is $\mathbf C(t,s)=\mathbf F^{-\mathsf T}(s)\mathbf C(t)\mathbf F^{-1}(s)$. A transient network's specific free energy is obtained by integrating these contributions over the formation history:
\begin{equation}
	\Psi_{\mathrm{TN}}(t)
	=
	\int_{-\infty}^{t}
	\xi(t,s)\,
	\psi_c\!\left(\mathbf F(t,s)\right)\,ds .
\end{equation}
For a compressible material, we define the chain population per unit mass, with the corresponding active chain density per unit formation volume given by $n_s(t,s)=\rho_s\xi(t,s)$. Since the material density satisfies $\rho_s J(s)=\rho_0$, the energy contribution per unit original reference volume is
\begin{equation}
J(s)n_s(t,s)\psi_c
=J(s)\rho_s\xi(t,s)\psi_c
=\rho_0\xi(t,s)\psi_c.    
\end{equation}
$J(s)$ therefore cancels through the density transformation. This differs from the engineering-energy convention adopted in our earlier work \cite{castonguay2026hereditary}, where each network is assigned a free-energy density per unit volume of its formation configuration, with parameters independent of the formation state, leading to $J(s)$-weighted history variables.

While conventional transient-network models typically assume a constant active chain population, we allow this population to evolve to accommodate processes such as irreversible damage and residual curing. To distinguish changes in population size from changes in the formation history, we define the total number of active chains per unit mass and the normalized formation-time distribution as
\begin{equation} \label{eqn:Tot_chains}
	N(t)=\int_{-\infty}^{t}\xi(t,s) \, ds,
\end{equation}
\begin{equation}
	\gamma(t,s)=\frac{\xi(t,s)}{N(t)},
	\qquad
	\int_{-\infty}^{t}\gamma(t,s)\,ds=1.
\end{equation}
Balanced breakage and reformation preserve $N(t)$, even as the network relaxes and its natural configuration evolves. 

We consider isochoric chain energies that depend on the relative deformation
through the first invariant. In this case,
\(\bar I_1(t,s)=\bar{\mathbf C}(t):\bar{\mathbf C}^{-1}(s)\), where
\(\bar{\mathbf C}(t)=J(t)^{-2/3}\mathbf C(t)\) and
\(\bar{\mathbf C}^{-1}(s)=J(s)^{2/3}\mathbf C^{-1}(s)\). Here, and
throughout this paper, the symbol $:$ denotes the Frobenius
inner product between tensors of the same order and double
contraction when a fourth-order tensor acts on a second-order
tensor. An overdot denotes a material time derivative at fixed reference
position, and for functions of both the current time $t$ and formation time $s$,
fixed $s$ as well. Expressed per unit
original reference volume, the isochoric energy takes the following form
\begin{equation}
	W_{\mathrm{iso}}(t)
	:=\rho_0 \Psi_{\mathrm{iso}}(t)=
	\rho_0 N(t)
	\int_{-\infty}^{t}
	\gamma(t,s)\,
	\psi_c\!\left(
	\bar{\mathbf C}(t):\bar{\mathbf C}^{-1}(s)
	\right)\,ds .
\end{equation}
Since \(\bar I_1(t,s)\) depends linearly on $\bar{\mathbf C}^{-1}(s)$, we define
its first history moment as
\begin{equation} \label{eqn:1st_histmoment}
	\mathbf A(t)
	=
	\int_{-\infty}^{t}
	\gamma(t,s)\,
	\bar{\mathbf C}^{-1}(s)\,ds .
\end{equation}
The corresponding reduced invariant is
\begin{equation}
	\bar Z(t)
	=
	\bar{\mathbf C}(t):\mathbf A(t)
	=
	\int_{-\infty}^{t}
	\gamma(t,s)\,
	\bar I_1(t,s)\,ds .
\end{equation}
The reduced model follows from a Taylor expansion of the chain energy
about the mean invariant $\bar Z$:
\begin{equation}
    \psi_c(\bar I_1)
    =
    \psi_c(\bar Z)
    +
    \psi_c'(\bar Z)(\bar I_1-\bar Z)
    +
    \frac{1}{2}\psi_c''(\bar Z)(\bar I_1-\bar Z)^2
    + \cdots .
\end{equation}
Since $\bar Z$ is the mean of the relative first invariant, the
linear term vanishes upon averaging. The leading correction is
therefore governed by its scalar variance,
\begin{equation}
    H(t)
    =
    \int_{-\infty}^{t}
    \gamma(t,s)
    \big(\bar I_1(t,s)-\bar Z(t)\big)^2\,ds .
    \label{eq:history_invariant_variance}
\end{equation}
To express this variance in terms of tensorial history variables,
we define the deviation of the inverse isochoric right
Cauchy--Green tensor from its mean:
\begin{equation}
    \mathbf X(t,s)
    =
    \bar{\mathbf C}^{-1}(s)-\mathbf A(t),
    \qquad
    \int_{-\infty}^{t}
    \gamma(t,s)\mathbf X(t,s)\,ds
    =\mathbf 0 .
\end{equation}
The corresponding fourth-order covariance tensor is
\begin{equation}
    \mathbf Q(t)
    =
    \int_{-\infty}^{t}
    \gamma(t,s)
    \mathbf X(t,s)\otimes\mathbf X(t,s)\,ds .
\end{equation}
It accounts for variability and correlations among the components
of the inverse isochoric right Cauchy--Green tensors. Since
$\bar I_1(t,s)-\bar Z(t)
=\bar{\mathbf C}(t):\mathbf X(t,s)$,
the scalar variance can be evaluated as
\begin{equation}
    H(t)
    =
    \bar{\mathbf C}(t):
    \big(\mathbf Q(t):\bar{\mathbf C}(t)\big).
\end{equation}
Substituting the expansion into the hereditary energy gives
\begin{equation}
    W_{\mathrm{iso}}(t)
    =
    \rho_0 N(t)
    \left[
        \psi_c(\bar Z)
        +
        \frac{1}{2}\psi_c''(\bar Z)
        \bar{\mathbf C}:(\mathbf Q:\bar{\mathbf C})
        + \cdots
    \right] .
\end{equation}
Higher-order approximations require progressively higher-rank history tensors, including, for example, sixth- and eighth-rank tensors associated with the third and fourth central moments (skewness and kurtosis), respectively.

Let $W_{\mathrm{iso}}^{(2)}$ refer to the energy retained through the variance correction.
The second Piola--Kirchhoff stress can be derived by differentiating $W_{\mathrm{iso}}^{(2)}$ with respect to the current deformation while holding $N$, $\mathbf{A}$, and $\mathbf{Q}$ fixed. Using the major symmetry of $\mathbf{Q}$ leads to the following auxiliary stress tensor
\begin{equation}
\tilde{\mathbf{S}}
	=
	2\frac{\partial W_{\mathrm{iso}}^{(2)}}
	{\partial\bar{\mathbf{C}}}
	=
	2\rho_0N
	\left[
	\left(
	\psi_c'(\bar{Z})
	+\frac{1}{2}\psi_c'''(\bar{Z})H
	\right)\mathbf{A}
	+
	\psi_c''(\bar{Z})
	\left(\mathbf{Q}:\bar{\mathbf{C}}\right)
	\right].
\end{equation}
The first-moment approximation is recovered by omitting the terms involving $H$ and $\mathbf{Q}$, leaving
$\tilde{\mathbf{S}}=2 \rho_0N\psi_c'(\bar{Z})\mathbf{A}$. For either approximation, the isochoric second Piola--Kirchhoff stress is
\begin{equation}
	\mathbf{S}_{\mathrm{iso}}
	=
	J^{-2/3}
	\left[
	\tilde{\mathbf{S}}
	-
	\frac{1}{3}
	\left(\tilde{\mathbf{S}}:\bar{\mathbf{C}}\right)
	\bar{\mathbf{C}}^{-1}
	\right].
\end{equation}
For numerical implementation, the consistent material tangent is obtained by further differentiating the stress, accounting for the dependence of the updated history variables on the current deformation.

While the isochoric first invariant case has been outlined here, the moment reduction also readily applies when the volumetric response is described by a transient network. Further details, including energies involving multiple relative invariants and the extension to anisotropy, can be found in \ref{app:multiple_invariants}.

\subsection{Network kinetics}
The detachment kinetics of all networks are assumed to be first-order, given by
\begin{equation} \label{eqn:kinetics}
	\begin{split}
		\dot{\xi}(t,s)=-k(t)  \xi(t,s) \quad \forall    t > s  \, 
	\end{split}
\end{equation}
where $k(t)$ represents the detachment rate. 
Equation \ref{eqn:kinetics} has the following general solution
\begin{equation} \label{eqn:gen_soln}
	\xi(t,s) =G(s)\exp\Big( -\intu_s^t k(\tau) d\tau \Big) \, .
\end{equation}
Here, $G(s)=\xi(s,s)$ is the chain generation rate at time $s$. 
Differentiating Equation \ref{eqn:Tot_chains} and using Equation \ref{eqn:kinetics} leads to $G(s)=\dot{N}(s)+ k(s)N(s)$.
Substituting into Equation \ref{eqn:gen_soln} and normalizing by $N(t)$ leads to
\begin{equation} \label{eqn:normalized_chain_solution}
	\gamma(t,s) =\Big(\frac{\dot{N}(s)}{N(t)}+ k(s)\frac{N(s)}{N(t)}\Big)\exp\Big( -\intu_s^t k(\tau) d\tau \Big) \, .
\end{equation}
With the kinetics defined, the moment-based history tensors can
be evaluated efficiently using the recurrence relations in
\ref{sec:Recurrence}. 

An example of an increasing active-chain population is presented in \ref{sec:Cure}, illustrating the effects of residual curing on the stress-free configuration.
Conversely, the population may decrease through irreversible chain loss. For the continuous evolution described here, the generation rate $G(t)=\dot N(t)+k(t)N(t)$ must remain nonnegative. Irreversible damage may be represented through additional chain detachment or abrupt population loss. In particular, uniform fractional removal of existing chains reduces $N$ without changing the normalized formation history or its moments. Developing a
damage evolution law and its coupling to network renewal is
outside the current scope.




\subsection{Arruda--Boyce chain energy}

The Gaussian chain statistics used in the classical transient network
model lead to a neo-Hookean response, which does not account for chain stiffening due to finite chain extensibility. Arruda and Boyce \cite{arruda1993three} incorporated
this behavior through an eight-chain representation of the network
based on Langevin chain statistics. In the present
isochoric formulation, the chain stretch is
$\lambda_{\mathrm{ch}}=\sqrt{\bar{I}_1/3}$, and the free energy per
chain can be written, up to a deformation-independent term, as
\begin{equation}
	\psi_c^{\mathrm{AB}}(\bar{I}_1)
	=
	N_c k_B T
	\left[
	x y + \ln\!\left(\frac{y}{\sinh y}\right)
	\right],
	\label{eq:AB_chain_energy}
\end{equation}
where
\begin{equation}
	x = \frac{\lambda_{\mathrm{ch}}}{\sqrt{N_c}}
	= \sqrt{\frac{\bar{I}_1}{3N_c}},
	\qquad
	y = \mathcal{L}^{-1}(x),
	\qquad
	\mathcal{L}(y) = \coth y - \frac{1}{y}.
	\label{eq:AB_Langevin}
\end{equation}
Here, $k_B$ is Boltzmann's constant, $T$ is the absolute temperature,
and $N_c$ is the number of statistical segments per chain, which
controls its extensibility. Full chain extension corresponds to
$x\to1$, or $\lambda_{\mathrm{ch}}\to\sqrt{N_c}$, at which the
inverse Langevin function diverges.

Using a series expansion of the inverse Langevin function,
Arruda and Boyce also expressed the strain energy
as a series in the first invariant and explicitly gave its first
five terms. Retaining these terms gives the polynomial approximation,
written here per chain and normalized to vanish at $\bar{I}_1=3$,
\begin{equation} \label{eqn:AB_5term}
	\begin{aligned}
		\psi_c^{\mathrm{AB5}}(\bar{I}_1)
		= k_B T \Bigg[
		&\frac{1}{2}(\bar{I}_1-3)
		+ \frac{1}{20N_c}(\bar{I}_1^2-9)
		+ \frac{11}{1050N_c^2}(\bar{I}_1^3-27)
		\\
		&+ \frac{19}{7000N_c^3}(\bar{I}_1^4-81)
		+ \frac{519}{673750N_c^4}(\bar{I}_1^5-243)
		\Bigg].
	\end{aligned}
\end{equation}
This polynomial form is widely used in commercial finite element
implementations, including Abaqus \cite{Abaqus2025} and LS-DYNA
\cite{LSDYNA2025}. 

\subsection{Connection to Lubliner’s model} \label{sec:Lubliner}


The moment reduction developed here also establishes a direct connection to Lubliner's internal-variable formulation \cite{Lubliner1985}. While our previous work \cite{castonguay2026hereditary} identified this correspondence for neo-Hookean elasticity, the present first-moment approximation extends it to nonlinear chain energies depending on the relative first invariant. Recall that retaining only the leading term gives
\begin{equation}
	W_{\mathrm{iso}}^{(1)}
	= \rho_0 N \psi_c(\bar{Z}),
	\qquad
	\bar{Z} = \bar{\mathbf{C}}:\mathbf{A}.
\end{equation}
For a constant active chain population, the normalized formation-history density satisfies $\dot{\gamma}(t,s)=-k(t)\gamma(t,s)$ and $\gamma(t,t)=k(t)$. Using these relations, and differentiating Equation \ref{eqn:1st_histmoment} leads to an equivalent rate form
\begin{equation}\label{eqn:1stmomentevolution}
	\dot{\mathbf{A}}
	= k(t)\left(\bar{\mathbf{C}}^{-1}-\mathbf{A}\right).
\end{equation}

The connection follows from the invariant structure of the multiplicative formulation. For $\mathbf F =\mathbf F_e \mathbf F_i$, with $\mathbf{C}_i=\mathbf{F}_i^{\mathsf{T}}\mathbf{F}_i$ and an isochoric inelastic deformation, the elastic isochoric first invariant is
\begin{equation}
	\bar{I}_{1e}
	= \operatorname{tr}\left(
	\mathbf{F}_i^{-\mathsf{T}}\bar{\mathbf{C}}\mathbf{F}_i^{-1}
	\right)
	= \bar{\mathbf{C}}:\mathbf{C}_i^{-1}.
\end{equation}
Thus, identifying $\mathbf{A}$ with $\mathbf{C}_i^{-1}$ formally recovers the same first-invariant dependence and linear evolution law as the corresponding isochoric Lubliner model.

The interpretation of this tensor is nevertheless different. An isochoric inelastic deformation requires $\det\mathbf{C}_i^{-1}=1$, which the linear evolution law does not generally preserve. In the present formulation, $\mathbf{A}$ is a normalized average of the inverse isochoric right Cauchy–Green tensors at formation. Although each $\bar{\mathbf{C}}^{-1}(s)$ has unit determinant, their average may not. Consequently, $\mathbf{A}$ remains a symmetric positive-definite history tensor without being constrained to represent a single isochoric inelastic deformation.

This interpretation distinguishes the moment approximation from the near-equilibrium linearization used to recover Lubliner's law within multiplicative formulations \cite{ReeseGovindjee1998FiniteViscoelasticity}. Here, the linear law governs the first history moment exactly under the assumed kinetics, whereas the approximation arises from evaluating the chain energy at the mean invariant. This is exact for neo-Hookean elasticity, while the variance term provides the next correction for nonlinear chain energies. Under isothermal renewal at constant active-chain population, \ref{app:dissipation} establishes dissipation of the first-moment model for nondecreasing chain energies and gives sufficient conditions for the second-moment model.


\section{Distributed network kinetics and spectral discretization}
\label{sec:distributed_network_kinetics}
The preceding formulation considers a single transient network population with a common detachment rate. 
To represent the non-exponential relaxation commonly observed in polymers, we extend the formulation by combining chain populations with different detachment rates, where each population retains its own history variables and recurrence relations. For constant detachment rates, their exponential survival functions combine to give a Prony-series representation, widely used to describe relaxation in viscoelastic constitutive models.

We first consider isothermal conditions, with each subpopulation maintaining a constant number of active chains. Let $w_i$ denote the fraction
of the total chain population assigned to rate $k_i$, where $w_i\geq0$,
$k_i>0$, and $\sum_{i=1}^{N_p}w_i=1$. The population balance developed in Section \ref{sec:TNTheory} then gives the combined normalized formation-history density
\begin{equation}
	\gamma(t,s)=\sum_{i=1}^{N_p}w_i k_i\exp[-k_i(t-s)].
	\label{eq:distributed_gamma}
\end{equation}
Writing the elapsed time since formation as $\tau=t-s$, we introduce the Prony series
\begin{equation}
	R(\tau)=\sum_{i=1}^{N_p}w_i\exp(-k_i\tau).
\end{equation}
This function gives the fraction of an initially present chain population that survives for a duration $\tau$. The normalized formation-history density is related to this survival function by
\begin{equation}
	\gamma(t,s)=-R'(t-s).
		\label{eq:gamma_survival_derivative}
\end{equation}
The discrete representation extends naturally to a continuous distribution of detachment rates, as follows
\begin{align}
	\gamma(t,s)&=\int_0^\infty k\exp[-k(t-s)]\rho(k)\,dk,
	\label{eq:continuous_gamma}\\
	R(\tau)&=\int_0^\infty \exp(-k\tau)\rho(k)\,dk,
	\qquad \rho(k)\geq0,\qquad\int_0^\infty\rho(k)\,dk=1.
	\label{eq:continuous_survival}
\end{align}
For ease of notation, we write the survival function as $R(t)$, with $t$ denoting elapsed time. We also refer to $R(t)$ as the relaxation function, since it describes normalized stress relaxation under ideal step-and-hold loading, as shown in Section~\ref{sec:calibration}. Unlike conventional hereditary representations in which the relaxation function weights the strain-rate history, the transient-network formulation uses its negative time derivative to weight the contributions of network populations formed at different times.



While the direct use of a discrete spectrum is the dominant approach in the literature and in practice, there are other suitable relaxation functions, including the stretched exponential \cite{husain2005modelling}, defined by 
\begin{equation} \label{eqn:KWW}
	R(t)= \exp(-(\alpha t)^{\beta}) \qquad \alpha>0,\quad 0<\beta < 1. 
\end{equation}	
For $\beta=1$, the expression reduces to a single exponential, while $\beta>1$ corresponds to a compressed exponential \cite{Ruta2013Compressed}.

Kohlrausch first introduced the stretched exponential to describe electrical discharge
\cite{kohlrausch1854theorie}, and Williams and Watts later applied it to
dielectric relaxation \cite{williams1970non}, giving rise to the name
Kohlrausch--Williams--Watts (KWW) function. Its subsequent use in describing viscoelastic response \cite{husain2005modelling}, structural relaxation \cite{phillips1996stretched}, fluorescence and luminescence decay \cite{lee2001application,chen2003apparent}, and magnetic relaxation \cite{chamberlin1984time} illustrates its broad applicability across diverse physical systems.

One widely used physical interpretation of KWW relaxation is that the non-exponential macroscopic response arises from a continuous distribution of independent first-order processes with different relaxation rates \cite{johnston2006stretched}.
Distributions of
activation barriers provide one physical origin of this
heterogeneity, as considered for biochemical transitions \cite{liebovitch1991distributions},
macromolecular relaxation \cite{edholm2000stretched}, and defect relaxation in disordered
solids
\cite{crandall1991defect}.
Related interpretations based on heterogeneous free-energy
landscapes have also been developed
\cite{wu2016heterogeneous}.
This interpretation connects naturally to transient networks, where
the individual processes correspond to chain populations with
different detachment rates, leading to the following equivalence:
\begin{equation}
	R(t)=\exp(-(\alpha t)^{\beta})	=	
	\int_{0}^{\infty}
	\rho(k)\exp(-kt)\,dk \, .
\end{equation}
Therefore, the rate density $\rho(k)$ is obtained by taking the inverse Laplace transform of the KWW function.

%


\subsection{Thermorheological processes and time-temperature superposition}
We first introduce temperature dependence through Arrhenius detachment rates. Allowing both the rate prefactor $\alpha_0$ and activation energy $E_a$ to vary among populations gives
\begin{equation}
	R(t,T)	= \int_{0}^{\infty} \int_{0}^{\infty} \rho(\alpha_0,E_a) \exp\left(-\alpha_0 \exp\left( \frac{-E_a}{\mathcal{R}T}\right) t\right) d\alpha_0 dE_a
\end{equation}
where  $\mathcal{R}$ is the universal gas constant and $\rho(\alpha_0,E_a)$ is a bi-variate probability distribution. 

For many polymers, the temperature dependence of relaxation admits a simpler description. Such materials are termed thermorheologically simple, meaning that their relaxation response at temperature $T$ can be obtained by shifting the response at a reference temperature $T_r$. This time–temperature superposition (TTS) relates the normalized relaxation functions through the shift factor $a_T$:
\begin{equation}
	R(t,T)=R\left( \frac{t}{a_T},T_r\right) \, .
\end{equation}	

There are two predominant shift factors used in practice, Arrhenius and Williams-Landel-Ferry (WLF)~\cite{williams1955temperature}, usually presented as follows
\begin{equation}
\text{Arrhenius: }	\ln (a_T)= \frac{-E_a}{\mathcal{R}} \left(\frac{1}{T_r}  -\frac{1}{T} \right) 
\end{equation}	
\begin{equation}
	\text{WLF: }\log_{10}(a_T)=-\frac{C_1(T-T_r)}{C_2+(T-T_r)}
\end{equation}
Both shift factors can be derived from the
Vogel--Fulcher--Tammann (VFT) form of the characteristic
relaxation rate \cite{Vogel1921,Fulcher1925,TammannHesse1926}:
\begin{equation} \label{eqn:VFT}
	 \alpha(T)=\alpha_0 \exp \Big(\frac{-E_a}{\mathcal{R}(T-C_T)}\Big)
\end{equation}
The shift factor follows by equating the dimensionless times
$\alpha(T)t$ and $\alpha(T_r)t/a_T$:
\begin{equation}
	a_T = \frac{\alpha(T_r)}{\alpha(T)}.
\end{equation}
Substituting the VFT rate law gives
\begin{equation}
	\ln (a_T)= \frac{-E_a}{\mathcal{R}} \left(\frac{1}{T_r-C_T}  -\frac{1}{T-C_T} \right) =\frac{-E_a}{\mathcal{R}} \frac{(T-T_r)}{(T_r-C_T)(T-C_T)}
\end{equation}	
The Arrhenius form is therefore recovered when the temperature offset $C_T$ is set to zero. Meanwhile, the WLF is recovered when 
\begin{equation}
	C_2 = T_r - C_T,
	\qquad
	C_1 = \frac{E_a}{\ln 10 \, \mathcal{R} \, C_2} \,.
\end{equation}
The equivalence between the VFT and WLF forms is well established \cite{dudowicz2015meaning}. Here, the VFT representation is adopted because it directly describes the temperature dependence of network detachment rates, while the corresponding WLF shift factors follow naturally from their relative temperature dependence.

For a continuous spectrum of linear processes, time--temperature
superposition with a VFT shift factor requires a common temperature
dependence across the rate spectrum. In this representation, the
populations share a single $E_a$ and temperature offset $C_T$,
while the prefactors $\alpha_0$ are distributed:
\begin{equation}
	R(t,T) = \int_0^{\infty} \rho(\alpha_0)
	\exp\left[
	-\alpha_0 \exp\left(-\frac{E_a}{\mathcal{R}(T-C_T)}\right)t
	\right] d\alpha_0 \, .
\end{equation}
This contrasts with the previously discussed interpretations of KWW relaxation based on
distributed activation energies
\cite{liebovitch1991distributions,edholm2000stretched}.

Thermorheological simplicity is not universal, as distinct relaxation mechanisms may exhibit different temperature dependencies, preventing their combined response from obeying time–temperature superposition \cite{Okamoto1991ModifiedStressOptical,Ngai1997ThermorheologicalComplexity}. As an example of modeling distinct relaxation contributions, sums of KWW functions have been used to describe the mechanical response of cortical bone \cite{Iyo2004Anisotropic,Iyo2006Mathematical}. Although the calibration considered here uses a single thermorheologically simple KWW process, multiple spectra with distinct temperature dependencies can be discretized and evolved independently using the same moment recurrence relations. This provides a straightforward extension to thermorheologically complex behavior arising from distinct relaxation mechanisms.

\FloatBarrier
\subsection{Calibration of thermally accelerated relaxation data} \label{sec:calibration}
Temperature-dependent viscoelastic relaxation can be characterized through a series of isothermal tests conducted at different temperatures. To relate the measured load retention (the force at a given time normalized by its initial value) to the network survival function, consider an ideal step-and-hold deformation under homogeneous uniaxial extension in the incompressible limit:
\begin{equation}
	\mathbf{F}(t)=
	\begin{cases}
		\mathbf{I}, & t<0,\\
		\operatorname{diag}(\lambda_0,\lambda_0^{-1/2},
		\lambda_0^{-1/2}), & t\geq0,
	\end{cases}
	\label{eq:relaxation_step_history}
\end{equation}
where $\lambda_0$ is the imposed stretch. The incompressible specialization
makes the relation to axial load explicit without introducing time-dependent
lateral deformation from a separate volumetric response. Alternatively, for a compressible material, the hyperelastic volumetric response can be replaced by a transient-network contribution governed by the same relaxation function as the isochoric response. Both stress contributions then relax proportionally during the hold, preserving the traction-free lateral condition and allowing the same calibration relation.

Here, we consider the full hereditary formulation of the transient network. All populations present before
the step share the original relaxed configuration and are subjected to the
same deformation. Populations formed after the step are formed stress-free in
the held configuration, so that
only surviving pre-step populations contribute to the axial stress during
the hold. For a uniaxial specimen with traction-free lateral
surfaces with constant populations and deformation-independent rates, the axial Cauchy stress $\sigma_{11}$ satisfies
\begin{equation}
	\sigma_{11}(t,T)
	=\sigma_{11}(0^+,T)R(t,T).
\end{equation}

Because the cross-sectional area remains constant during the hold,
the normalized axial force equals the normalized axial stress.
The load retention is therefore
\begin{equation}
	LR(t,T)
	=\frac{\sigma_{11}(t,T)}{\sigma_{11}(0^+,T)}
	=R(t,T).
\end{equation}
We fit the KWW response in Equation \ref{eqn:KWW}, using the temperature-dependent
rate scale in Equation \ref{eqn:VFT}, simultaneously to measurements at all
temperatures. No preliminary shifting to a reference temperature
is required.
Given measured load-retention values
$L_{n,j}=LR(t_{n,j},T_j)$ at times $t_{n,j}$ and temperatures $T_j$,
we determine $\boldsymbol{p}=(\alpha_0,E_a,C_T,\beta)$ by minimizing
\begin{equation}
	\widehat{\boldsymbol{p}}
	=\underset{\boldsymbol{p}}{\operatorname{arg\,min}}
	\sum_{j=1}^{N_T}\sum_{n=1}^{N_j}
	\left\{
	\exp\!\left[-\bigl(\alpha(T_j;\boldsymbol{p})t_{n,j}\bigr)^\beta\right]
	-L_{n,j}
	\right\}^{2}.
	\label{eq:kww_joint_fit}
\end{equation}

The admissible parameters satisfy $\alpha_0>0$, $E_a>0$,
$0<\beta<1$, and $C_T<\min_j T_j$ (to remain above the
VFT singular temperature). For an Arrhenius fit,
$C_T$ is fixed to zero, leaving three unknowns.

\subsection{Evaluation and discretization of the rate spectrum}

Once the kinetic parameters have been identified, the continuous
spectrum is discretized into a finite set of first-order processes.
Introducing the dimensionless time $\theta=\alpha(T)t$ and rate
$x=k/\alpha(T)$ gives
\begin{equation}
	\exp(-\theta^\beta)
	=\int_0^\infty \rho_\beta(x)\exp(-x\theta)\,dx.
	\label{eq:dimensionless_kww_spectrum}
\end{equation}
The normalized density $\rho_\beta(x)$ depends only on the stretching exponent $\beta$. For $0<\beta<1$, it is the inverse Laplace transform of the KWW function and corresponds to a one-sided positive Lévy-stable density \cite{jurlewicz1993relationship,elton2018stretched}. Although simple closed-form expressions are available only for special values of $\beta$, the density can be evaluated using integral and series representations derived from Bromwich inversion \cite{johnston2006stretched,elton2018stretched}.

Previous approximations of the KWW function by finite sums of exponentials include discretization of a truncated rate spectrum using equal-width intervals \cite{Anderssen2011Kohlrausch} and optimization of discrete Prony coefficients and rates \cite{Mauro2018Prony}. Here, we instead partition the continuous rate spectrum into bins of equal probability, using its cumulative distribution and inverse,
\begin{equation}
	F_\beta(x)=\int_0^x \rho_\beta(u)\,du,
	\qquad
	F_\beta^{-1}(q), \quad 0<q<1.
\end{equation}
These quantities are evaluated using \texttt{scipy.stats.levy\_stable} \cite{virtanen2020scipy,scipyLevyStableDocs}, which employs Nolan’s piecewise integration method \cite{nolan1997numerical} for evaluating the cumulative distribution function. Stable distributions admit several parameter conventions, of which we use SciPy's default $S_1$ convention, where $\rho_\beta$ has stability index $\beta$, skewness parameter one, location zero, and scale
\begin{equation}
	c_\beta=\left[\cos\left(\frac{\pi\beta}{2}\right)\right]^{1/\beta}.
\end{equation}

The continuous spectrum is partitioned into $N_p$ bins of equal
probability, each represented by a discrete rate at its median. The resulting dimensionless rates and weights are
\begin{equation}
	x_i=F_\beta^{-1}\!\left(\frac{i-\tfrac12}{N_p}\right),
	\qquad
	w_i=\frac{1}{N_p},
	\qquad i=1,\ldots,N_p.
	\label{eq:kww_quantile_discretization}
\end{equation}
\Needspace{4\baselineskip}
For \(0<\beta<1\), the continuous KWW function has a singular initial slope, which cannot be reproduced by a finite Prony series \cite{Mauro2018Prony}. The discretization therefore regularizes the short-time response by introducing a finite initial relaxation rate.

This approach separates the calibration of the continuous relaxation spectrum from its discrete representation. Once the KWW parameters have been identified, the number of Prony terms can be adjusted to balance approximation accuracy against computational cost without refitting the experimental data. Furthermore, the discretization produces positive, normalized weights without the potentially ill-conditioned optimization associated with independently fitting Prony coefficients and relaxation rates. These advantages motivate the use of equal-probability binning here, although alternative discretization strategies may further reduce the approximation error for a fixed number of terms.


%


\section{Examples}
\label{sec:examples}
The following examples assess two distinct approximations introduced in this work: the reduction of the hereditary deformation history to a finite set of tensorial moments, and the discretization of a continuous spectrum of detachment rates. The first set of examples examines whether retaining the second moment provides a meaningful improvement over the first-moment approximation, particularly under complex loading histories. The second evaluates how accurately a calibrated, temperature-dependent KWW relaxation response can be represented by a finite Prony series, enabling efficient numerical implementation.

\subsection{Three-term Arruda--Boyce moment approximation}
\label{sec:ab3_benchmarks}

To isolate the error introduced by the moment approximation, we use
the first three terms of the Arruda--Boyce (AB3) energy in Equation \ref{eqn:AB_5term}.
Because this energy is cubic in the relative first invariant, its
hereditary average can be evaluated exactly using raw history tensors
of ranks two, four, and six, following the recurrence construction
developed in our earlier work \cite{castonguay2026hereditary}. It therefore provides a reference free of
moment-truncation error against which we assess the first- and
second-moment approximations.

All four benchmarks are homogeneous, deformation-controlled calculations
using the same time-integration scheme for each moment order.
A single transient population with constant active chain density
and temperature is considered. The material is initially relaxed, consistent with a stationary configuration prior to loading, such that $\mathbf F(0)=\mathbf I$, $\mathbf A(0)=\mathbf I$, and $\mathbf Q(0)=\mathbf 0$. For the three-term Arruda–Boyce energy, the instantaneous infinitesimal shear modulus of the initially relaxed network is
\begin{equation}
	\mu = \rho_0 N k_B T
	\left(1+\frac{3}{5N_c}+\frac{99}{175N_c^2}\right).
\end{equation}
For simplicity, we choose stress and time units such that
$\mu=1$ and $k=1$. The remaining parameters are
\begin{equation}
	K_{\mathrm{bulk}}=100,
	\qquad N_c=5,
	\qquad \Delta t=0.001 \, ,
\end{equation}
where for the compressible case, the volumetric energy is taken to be
\begin{equation}
	W_{\text{vol}}(J)
	=
	\frac{K_{\mathrm{bulk}}}{2}
	\left[
	\frac{1}{2}(J^2-1)-\ln J
	\right] \, .
\end{equation}
We consider the following four loading histories:

\begin{enumerate}
	\item In the uniaxial ramp--hold--unload history,
	\(\mathbf F=\operatorname{diag}(\lambda,\lambda^{-1/2},\lambda^{-1/2})\).
	The stretch increases linearly from \(1\) to \(2\) over
	\(0\leq t\leq1\), is held until \(t=5\), returns linearly to one by
	\(t=6\), and is then held until \(t=10\).
	
	\item In cyclic simple shear,
	\(\mathbf F=\mathbf I+\gamma\mathbf e_1\otimes\mathbf e_2\).
	The shear history varies linearly between the points
	\((t,\gamma)=(0,0),(1,1.5),(3,-1.5),(5,1.5),(7,-1.5),(8,0)\).
	
	\item The non-proportional history uses
	\begin{equation}
		\mathbf F
		=
		\begin{bmatrix}
			1 & \gamma_{12} & 0\\
			0 & 1 & \gamma_{23}\\
			0 & 0 & 1
		\end{bmatrix}.
	\end{equation}
	Here \(\gamma_{12}\) ramps from zero to \(1.1\) over
	\(0\leq t\leq1\) and is held while \(\gamma_{23}\) ramps from zero to
	\(1.1\) over \(3\leq t\leq4\).  Both shears are held until \(t=6\),
	after which \(\gamma_{12}\) reverses to \(-1.1\) by \(t=7\), with
	\(\gamma_{23}=1.1\) held through \(t=10\).
	
	\item The compressible uniaxial history combines an isochoric axial
	stretch with a volume change,
	\begin{equation}
		\mathbf F
		=
		J^{1/3}
		\operatorname{diag}(\lambda,\lambda^{-1/2},\lambda^{-1/2}).
	\end{equation}
	Both \(\lambda\) and \(J\) increase linearly from \(1\) to \(1.8\) and
	\(1.1\), respectively, over \(0\leq t\leq1\), are held until \(t=5\),
	return to one by \(t=6\), and are then held until \(t=10\).
\end{enumerate}
For approximation order \(m\), the reported stress error is
\begin{equation}
	e_m(t)
	=
	\frac{
		\lVert
		\operatorname{dev}\boldsymbol\sigma^{(m)}(t)
		-
		\operatorname{dev}\boldsymbol\sigma^{\mathrm{AB3}}(t)
		\rVert_F
	}{
		\displaystyle\max_{0\leq s\leq t_{\mathrm{end}}}
		\lVert
		\operatorname{dev}\boldsymbol\sigma^{\mathrm{AB3}}(s)
		\rVert_F
	},
	\label{eq:normalized_stress_error}
\end{equation}
where \(\lVert\cdot\rVert_F\) is the Frobenius norm and the denominator is
the maximum norm over the complete exact AB3 history.  
All stress errors are evaluated using the deviatoric Cauchy stress. This coincides with the total stress for the three volume-preserving histories, while isolating the transient-network response in the compressible case. 

\begin{figure}[]
	\centering
	\includegraphics[width=0.80\linewidth]{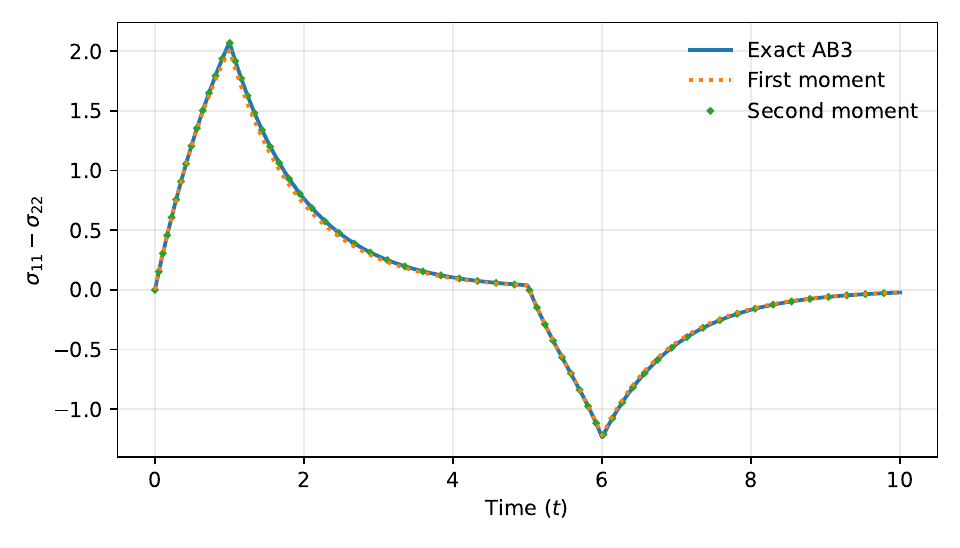}
	\caption{Axial stress difference $\sigma_{11}-\sigma_{22}$ from a volume-preserving uniaxial ramp, hold, unload, and final hold at the original configuration.}
	\label{fig:ab3_uniaxial}
\end{figure}
\begin{figure}[]
	\centering
	\includegraphics[width=0.8\linewidth]{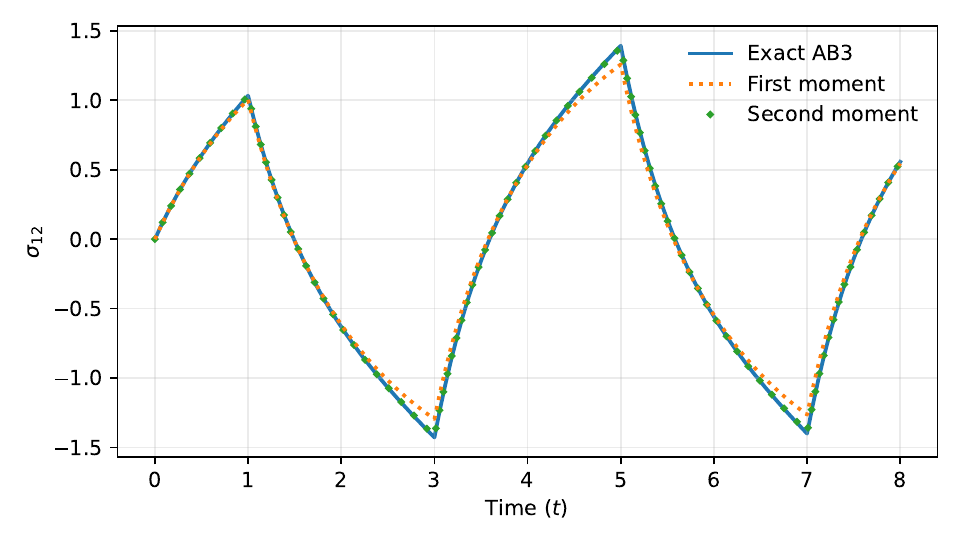}
	\caption{Shear stress $\sigma_{12}$ from cyclic simple shear.}
	\label{fig:ab3_cyclic}
\end{figure}
\begin{figure}[]
	\centering
	\includegraphics[width=0.8\linewidth]{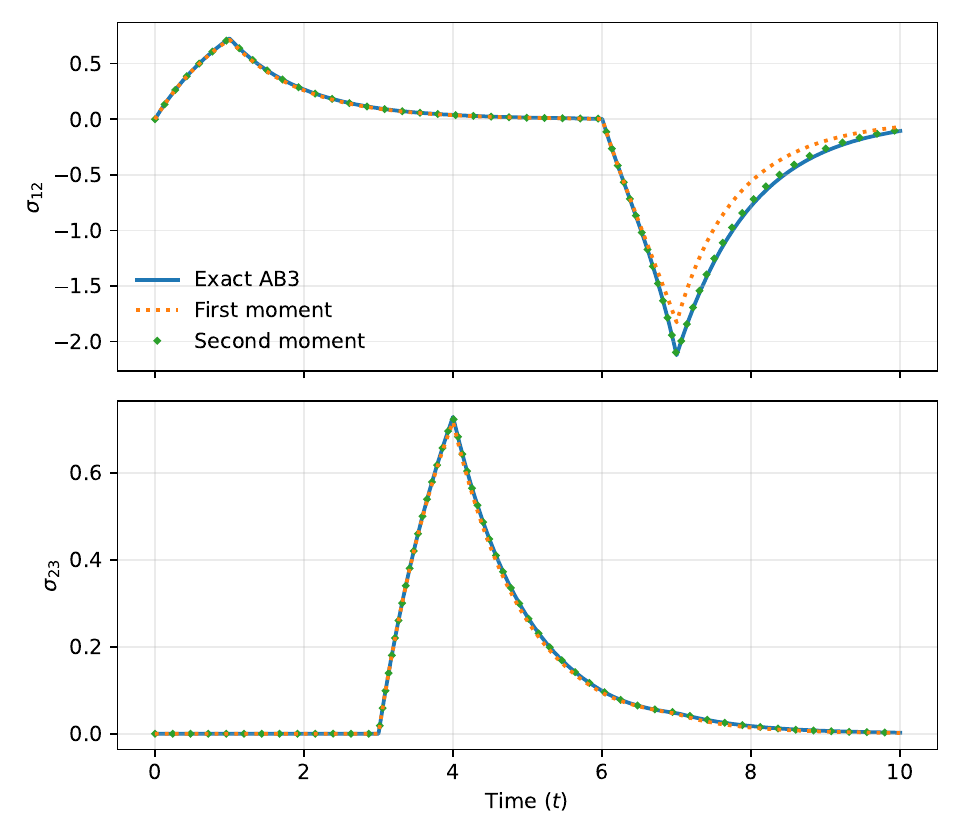}
	\caption{Shear stresses $\sigma_{12}$ and $\sigma_{23}$ from non-proportional sequential shear.}
	\label{fig:ab3_nonproportional}
\end{figure}
\begin{figure}[]
	\centering
	\includegraphics[width=0.8\linewidth]{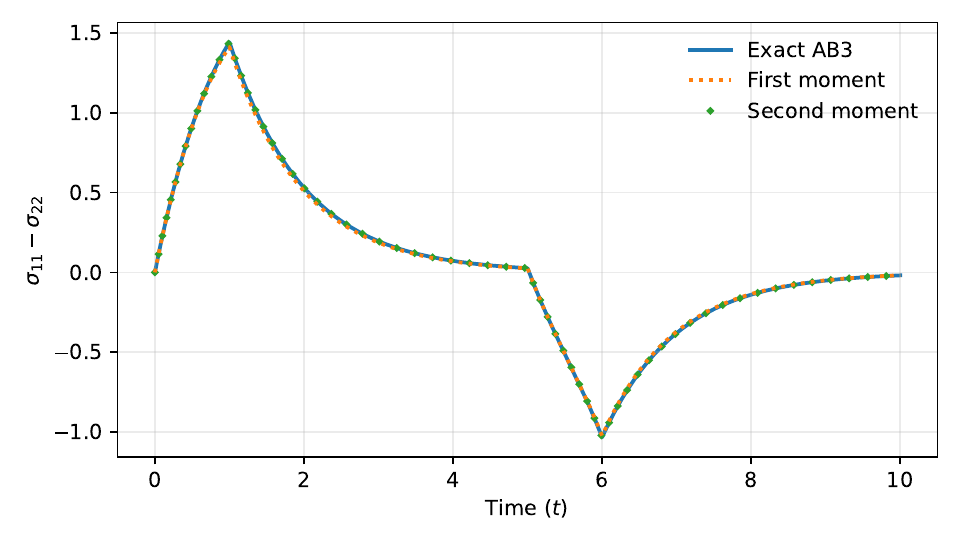}
	\caption{Axial stress difference $\sigma_{11}-\sigma_{22}$ from a combined volumetric and isochoric uniaxial ramp, hold, unload, and final hold at the original configuration. }
	\label{fig:ab3_compressible}
\end{figure}

Figures~\ref{fig:ab3_uniaxial} -- \ref{fig:ab3_compressible} show selected stress components, while the errors in Table~\ref{tab:ab3_errors} are evaluated using the complete deviatoric stress tensor.
Peak errors decrease from 2.16--17.78\% for the first-moment
approximation to 0.0508--2.04\% for the second-moment approximation.
The cyclic and non-proportional histories remain the most demanding,
as load reversals and changes in deformation plane produce more 
complex distributions of network formation configurations.

Halving the time step changes 
the first- and second-moment maximum errors by less than $4.5\times10^{-5}$ and $5.2\times10^{-6}$, respectively, across all four histories.
These results demonstrate that the first-moment approximation captures the overall stress evolution, while inclusion of the second moment substantially improves quantitative agreement with the exact hereditary response, particularly under cyclic and non-proportional loading.

\begin{table}[]	
	\centering
	\begin{tabular}{@{}lrrrr@{}}
		\toprule
		& \multicolumn{2}{c}{First moment} & \multicolumn{2}{c}{Second moment}\\
		\cmidrule(lr){2-3}\cmidrule(l){4-5}
		History & Maximum (\%) & RMS (\%) & Maximum (\%) & RMS (\%)\\
		\midrule
		Uniaxial            & 3.35  & 1.20 & 0.112 & 0.0442\\
		Cyclic shear        & 12.60 & 4.83 & 1.18 & 0.418\\
		Non-proportional    & 17.78 & 5.90 & 2.04 & 0.781\\
		Compressible uniaxial & 2.16 & 0.812 & 0.0508 & 0.0203\\
		\bottomrule
	\end{tabular}
		\caption{Maximum and RMS values of the normalized deviatoric stress error $e_m(t)$ defined in Equation~\ref{eq:normalized_stress_error}, expressed as percentages.}
	\label{tab:ab3_errors}	
\end{table}

\FloatBarrier

\subsection{Thermally accelerated relaxation of Butyl-A}
\label{sec:butyl_a_fit}

We next illustrate the identification and discretization of distributed network
kinetics using the thermally accelerated relaxation measurements of
Gillen et al.~\cite{Gillen2003Validation}. Their experiments measured the sealing force of lubricated butyl rubber
O-ring segments (Butyl-A) compressed by approximately 25\% during
aging in air. Force measurements were taken approximately isothermally
at the aging temperature and normalized by an extrapolated initial
value $F_0$. We use 66 measurements digitized from their Fig.~7
at $80$, $95$, and $110\,^{\circ}\mathrm{C}$, retaining both reported series at 95 and \(110^\circ\mathrm C\) as separate observations. The $125\,^{\circ}\mathrm{C}$ data
are excluded because Gillen et al.\ suspected diffusion-limited
oxidation above $110\,^{\circ}\mathrm{C}$. Here, we use these measurements to demonstrate calibration of an effective temperature-dependent relaxation spectrum, without attempting to distinguish the mechanisms responsible for the observed force decay.

Introducing a VFT temperature offset did not materially improve
the fit, so we adopt the KWW-Arrhenius model ($C_T=0$).
Its parameters $\alpha_0$, $E_a$, and $\beta$ are identified
simultaneously from the unshifted measurements at all three
temperatures by unweighted least squares, with time expressed
in weeks.

\begin{figure}[h]
	\centering
	\includegraphics[width=0.9\linewidth]{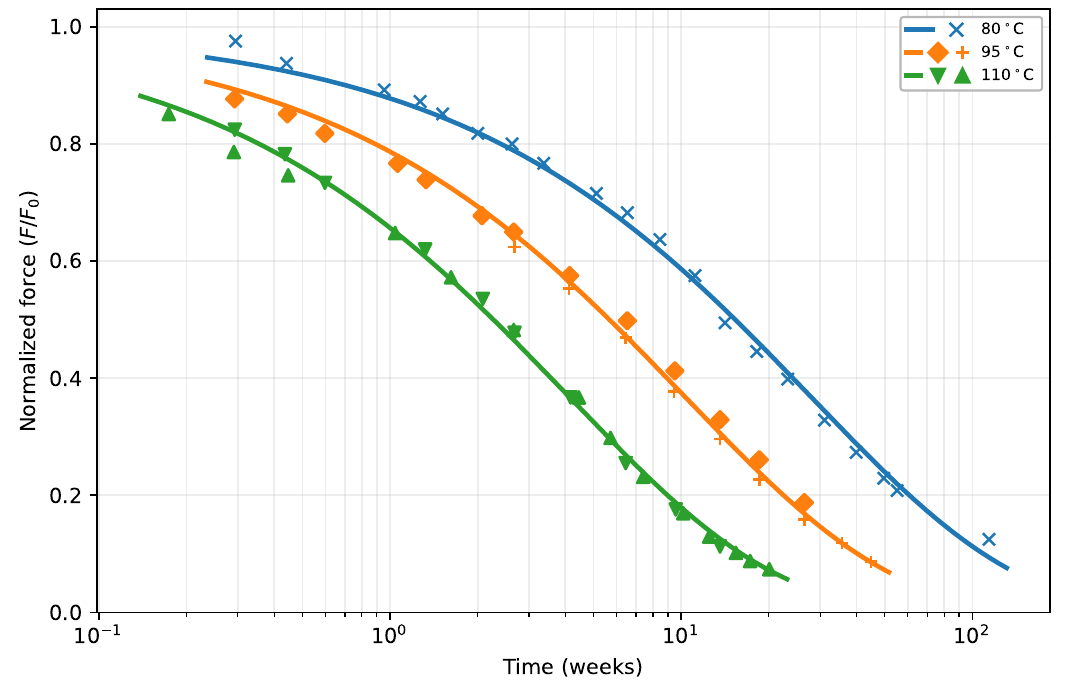}
	\caption{Simultaneous KWW--Arrhenius fit to the thermally accelerated
		Butyl-A relaxation measurements. Distinct symbols identify the five
		selected experimental marker series digitized from
		Gillen et al.~\cite{Gillen2003Validation}; colored solid lines use one common stretching exponent and an Arrhenius rate scale.}
	\label{fig:butyl_a_kww_fit_main}
\end{figure}

\begin{table}[h]
	\centering
	\begin{tabular}{@{}lc@{}}
		\toprule
		Parameter & Estimate\\
		\midrule
$\alpha_0$ ($\mathrm{week}^{-1}$) & $1.46\times10^{9}$ \\
$E_a$ ($\mathrm{kJ\,mol^{-1}}$)   & $71.73$ \\
$\beta$                         & $0.6118$ \\
		\bottomrule
	\end{tabular}
	\caption{Parameters of the simultaneous KWW--Arrhenius fit
		to the digitized Butyl-A relaxation measurements.}
	\label{tab:butyl_a_fit_main}
\end{table}
\Needspace{5\baselineskip}
Figure~\ref{fig:butyl_a_kww_fit_main} shows good agreement between the fitted KWW--Arrhenius response and the measurements. The root-mean-square and maximum absolute residuals are 0.01515 and 0.03543, respectively, in normalized force. The fitted activation energy of $71.73 \, \mathrm{kJ\,mol^{-1}}$ is reasonably close to the approximately $78 \, \mathrm{kJ\,mol^{-1}}$ obtained by Gillen et al.\ through empirical shifting of individual relaxation curves.

\subsubsection{Prony discretization of the fitted spectrum}
\label{sec:prony_term_example}
To evaluate the Prony approximation over the range covered by the measurements, we map the earliest and latest observations at each temperature to dimensionless time $\theta=\alpha(T)t$, giving
\begin{equation}
	\theta_{\min}=0.0105682,
	\qquad
	\theta_{\max}=4.87268.
\end{equation}
We discretize the fitted spectrum using Equation \ref{eq:kww_quantile_discretization} with $N_p = 4, 8, 16,$ and $32$ Prony terms, keeping the fitted parameters
fixed. The measurement interval is used only to evaluate the approximation error, not to determine the discrete rates and weights.

\begin{figure}[htbp]
	\centering
	\includegraphics[width=0.86\linewidth]{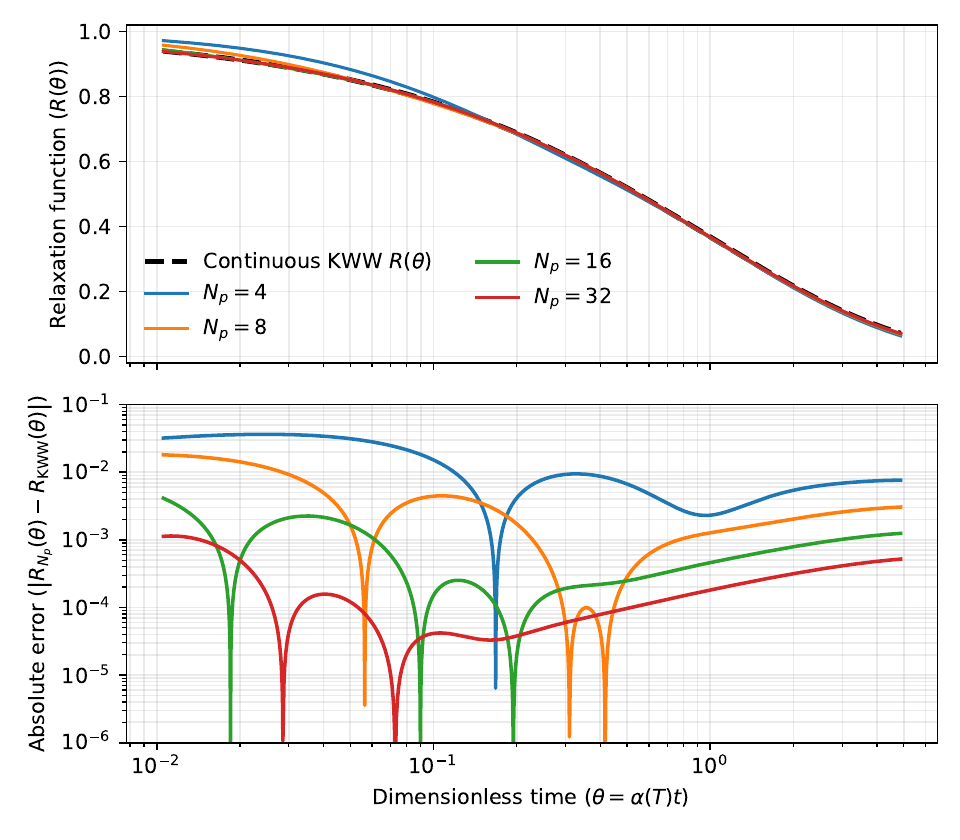}
	\caption{Prony approximations of the fitted Butyl-A KWW
		scalar relaxation function $R(\theta)$ over the dimensionless-time interval spanned by
		the selected measurements. The lower panel shows the absolute
		difference from the continuous KWW function.}
	\label{fig:prony_term_convergence}
\end{figure}

Figure~\ref{fig:prony_term_convergence} shows the resulting relaxation
functions and their absolute differences from the KWW fit. For the
approximations considered, increasing the number of terms reduces the
maximum and RMS errors over the selected interval. To evaluate the Prony approximation across widely separated relaxation times without disproportionately weighting the later-time response, we define the RMS error over logarithmic time:
\begin{equation}
	e_{\mathrm{RMS}}
	=
	\left[
	\frac{1}{\ln(\theta_{\max}/\theta_{\min})}
	\int_{\theta_{\min}}^{\theta_{\max}}
	\left(R_{N_p}(\theta)-R_{\mathrm{KWW}}(\theta)\right)^2
	\frac{d\theta}{\theta}
	\right]^{1/2}.
\end{equation}
The maximum error is the largest absolute difference over the
same interval. The maximum and RMS errors are reported in Table~\ref{tab:prony_term_errors}.

\begin{table}[htbp]
	\centering
	\begin{tabular}{@{}rrr@{}}
		\toprule
		Total terms $N_p$ & Maximum error (\%) & RMS error (\%)\\
		\midrule
		4  & 3.64  & 2.01\\
		8  & 1.81  & 0.671\\
		16 & 0.413 & 0.116\\
		32 & 0.114 & 0.0369\\
		\bottomrule
	\end{tabular}
\caption{Errors in the Prony approximation of the fitted Butyl-A
	KWW relaxation function, expressed as percentages of the initial
	response $R(0)=1$.}	\label{tab:prony_term_errors}
\end{table}

\section{Summary}
The moment reduction developed here retains the hereditary interpretation of transient-network theory while replacing the full formation history with a finite set of tensorial moments. In particular, the first-moment approximation recovers the corresponding Lubliner model and provides a physical interpretation of its internal variable as the mean inverse formation metric, rather than the inverse metric associated with a single inelastic deformation. This distinction avoids imposing the isochoric constraint of a multiplicative decomposition on the averaged formation history. For nonlinear chain elasticity, higher moments provide a systematic means of accounting for the distribution of network formation configurations.

The numerical comparisons with the exact hereditary response demonstrate the importance of these higher moments under complex loading histories. Retaining the second moment reduces the largest peak stress error from approximately 18\% to 2\% across the examples considered, with particularly significant improvements under cyclic and non-proportional loading. However, the accuracy of a truncated moment expansion necessarily depends on the distribution of formation configurations and the nonlinearity of the chain energy. Furthermore, under isothermal conditions with a constant active-chain population, the first-moment approximation is dissipative for any nondecreasing chain energy. The second-moment approximation is also dissipative for the three-term Arruda–Boyce energy considered here, although this is not guaranteed for arbitrary chain energies.

Distributed detachment rates can be incorporated into the same recursive moment framework, allowing non-exponential relaxation to be represented efficiently. By calibrating the continuous KWW response before discretizing its rate spectrum, the number of Prony terms can be adjusted without refitting the experimental data. Together, these developments provide a practical route from hereditary integral transient-network theory to finite-element implementation, separating the approximation of deformation history from the discretization of distributed kinetics while retaining a direct connection to chain elasticity, network renewal, and temperature-dependent relaxation.

\section*{Acknowledgement}
This work was performed under the auspices of the U.S. Department of Energy by Lawrence Livermore National Laboratory under the Contract DE-AC52-07NA27344, Release number LLNL-JRNL-2026117.

\section*{AI-Assisted Tools Disclosure}
Codex was used to assist with code implementation and the setup of numerical calculations. ChatGPT was used for error checking and improving the manuscript’s organization and flow. The authors verified the code and results and take full responsibility for the manuscript’s contents.

\appendix
\counterwithin*{figure}{section}

\section{Recurrence relations for history tensors} \label{sec:Recurrence}
Consider a time step from $t_N$ to $t_{N+1}$, with
$\Delta t=t_{N+1}-t_N$. The normalized chain density at the end
of the step is
\begin{equation}
	\label{eqn:chain_solution_endpoint}
	\gamma(t_{N+1},s)
	=
	\frac{\dot N(s)+k(s)N(s)}{N(t_{N+1})}
	\exp\left(
	-\int_s^{t_{N+1}}k(\tau)\,d\tau
	\right).
\end{equation}
Taking the detachment rate to be constant within the increment,
$k(\tau)=\tilde k$, define the fraction of the updated population
consisting of chains that survive from $t_N$ as
\begin{equation}
	r_N
	=
	\frac{N(t_N)}{N(t_{N+1})}
	\exp(-\tilde k\Delta t).
\end{equation}
For populations already present at $t_N$, the chain density
therefore satisfies
\begin{equation}
	\gamma(t_{N+1},s)=r_N\gamma(t_N,s).
\end{equation}
Splitting the first history moment into surviving and newly
formed populations gives
\begin{equation}
	\begin{aligned}
		\mathbf A(t_{N+1})
		&=
		\int_{-\infty}^{t_{N+1}}
		\gamma(t_{N+1},s)\,
		\bar{\mathbf C}^{-1}(s)\,ds
		\\
		&=
		r_N\mathbf A(t_N)
		+
		\int_{t_N}^{t_{N+1}}
		\gamma(t_{N+1},s)\,
		\bar{\mathbf C}^{-1}(s)\,ds.
	\end{aligned}
\end{equation}
Normalization of $\gamma$ implies that the newly formed
population has total weight
\begin{equation}
	\int_{t_N}^{t_{N+1}}\gamma(t_{N+1},s)\,ds
	=
	1-r_N.
\end{equation}
Approximating the formation configurations within the increment
by the configuration at $t_{N+1}$ then gives the discrete update
\begin{equation}
	\mathbf A(t_{N+1})
	=
	r_N\mathbf A(t_N)
	+
	(1-r_N)\bar{\mathbf C}^{-1}(t_{N+1}).
\end{equation}
The recurrence extends directly to higher raw moments.
The raw second history moment is the fourth-order tensor
\begin{equation}
	\label{eq:raw_second_history_moment}
	\mathbf M(t)
	=
	\int_{-\infty}^{t}
	\gamma(t,s)\,
	\bar{\mathbf C}^{-1}(s)
	\otimes
	\bar{\mathbf C}^{-1}(s)\,ds.
\end{equation}
Using the same endpoint approximation gives
\begin{equation}
	\label{eq:raw_second_history_update}
	\begin{aligned}
		\mathbf M(t_{N+1})
		&=
		r_N\mathbf M(t_N)+
		(1-r_N)
		\bar{\mathbf C}^{-1}(t_{N+1})
		\otimes
		\bar{\mathbf C}^{-1}(t_{N+1}).
	\end{aligned}
\end{equation}
After updating the raw moments, the covariance tensor and the
scalar variance of the relative first invariant are recovered as
\begin{equation}
	\begin{aligned}
		\mathbf Q
		&=
		\mathbf M-\mathbf A\otimes\mathbf A,
		\\
		H
		&=
		\bar{\mathbf C}:
		\left[
		(\mathbf M-\mathbf A\otimes\mathbf A):
		\bar{\mathbf C}
		\right].
	\end{aligned}
\end{equation}
Here, all quantities are evaluated at $t_{N+1}$. Higher raw moments follow the same
recurrence, with endpoint contributions formed from the
corresponding higher-order tensor products of
$\bar{\mathbf C}^{-1}(t_{N+1})$.
\section{Dissipation of the moment approximations}
\label{app:dissipation}
Consider isothermal conditions, a constant active-chain
population $N$, and deformation-independent renewal at rate
$k(t)\geq 0$. For approximation order $m=1,2$, the Helmholtz
free energy per reference volume depends on the current
deformation and retained history tensors:
\begin{equation}
	\begin{aligned}
		W^{(m)}(\mathbf C,\mathbf A,\mathbf Q)
		&=
		W_{\mathrm{vol}}(J)+
		W_{\mathrm{iso}}^{(m)}
		(\bar{\mathbf C},\mathbf A,\mathbf Q).
	\end{aligned}
\end{equation}
The first-moment approximation is independent of $\mathbf Q$.

Under these assumptions, the mechanical dissipation inequality is
\begin{equation}
	\mathcal D^{(m)}
	=
	\frac12\mathbf S^{(m)}:\dot{\mathbf C}
	-
	\dot W^{(m)}
	\geq 0,
\end{equation}
where $\mathbf S^{(m)}$ is the total second Piola--Kirchhoff
stress. Applying the chain rule gives
\begin{equation}
		\dot W^{(m)}=
		\frac{\partial W^{(m)}}{\partial\mathbf C}
		:\dot{\mathbf C}
		+
		\frac{\partial W^{(m)}}{\partial\mathbf A}
		:\dot{\mathbf A}
		+
		\frac{\partial W^{(m)}}{\partial\mathbf Q} : \dot{\mathbf Q}	\,	.
\end{equation}
The stress is obtained by differentiating the energy with respect
to the current deformation while holding the history tensors fixed:
\begin{equation}
	\mathbf S^{(m)}
	=
	2\left.
	\frac{\partial W^{(m)}}{\partial\mathbf C}
	\right|_{\mathbf A,\mathbf Q}.
\end{equation}
The deformation-rate contribution to $\dot W^{(m)}$ therefore
cancels against the stress power, leaving
\begin{equation}
		\mathcal D^{(m)}
		= -	\frac{\partial W_{\mathrm{iso}}^{(m)}}{\partial\mathbf A}
		:\dot{\mathbf A}
		-	\frac{\partial W_{\mathrm{iso}}^{(m)}}{\partial\mathbf Q}:
		\dot{\mathbf Q}=
		-\left.\dot W_{\mathrm{iso}}^{(m)}\right|_{\mathbf C}.
\end{equation}
The notation $\left.\dot{(\cdot)}\right|_{\mathbf C}$ denotes
the contribution to the material time derivative from
history-tensor evolution, evaluated with the current
deformation held fixed.

In Section \ref{sec:Lubliner}, an explicit rate form is presented in Equation \ref{eqn:1stmomentevolution} for $\dot{\mathbf A}$.
A similar rate form can be derived for $\mathbf Q$ as well.
Using the Leibniz rule, and exploiting $\dot{\mathbf X}=-\dot{\mathbf A}$ and $\int_{-\infty}^{t} \gamma(t,s) 	\mathbf X(t,s)\, ds =0$, we have
\begin{equation}
	\dot{\mathbf Q}
	=
	\gamma(t,t)	\mathbf X(t,t)\otimes\mathbf X(t,t) +
	\int_{-\infty}^{t}
	\dot{\gamma}(t,s)\,
	\mathbf X(t,s)\otimes\mathbf X(t,s)\,ds .
\end{equation}
Using \(\gamma(t,t)=k(t)\) and \(\dot\gamma(t,s)=-k(t)\gamma(t,s)\) leads to  
\begin{equation}
	\dot{\mathbf Q}
	=
	k(t)\Big( \mathbf X(t,t)\otimes\mathbf X(t,t) - \mathbf Q \Big) \, .
\end{equation}
Now, the relevant partial derivatives are given by
\begin{align} \label{eqn:partialderiv}
	\frac{\partial W_{\mathrm{iso}}^{(2)}}{\partial \mathbf{A}}
	&=
	\rho_0 N
	\left[
	\psi_c'(\bar{Z})
	+\frac{1}{2}\psi_c'''(\bar{Z})H
	\right]\bar{\mathbf{C}},
	\\
	\frac{\partial W_{\mathrm{iso}}^{(2)}}{\partial \mathbf{Q}}
	&=
	\frac{\rho_0 N}{2}\psi_c''(\bar{Z})
	\,\bar{\mathbf{C}}\otimes\bar{\mathbf{C}}.
\end{align}
Then note that
\begin{equation}
 \bar{\mathbf{C}}:(\bar{\mathbf{C}}^{-1} -\mathbf{A})=-a \, , 
\end{equation}
\begin{equation}
(\bar{\mathbf{C}}\otimes\bar{\mathbf{C}}):\bigl[\mathbf X(t,t)\otimes\mathbf X(t,t)- \mathbf Q \bigr] \, 
=a^{2}-H 
\end{equation}
where $a=\bar Z-3$. Since each relative isochoric invariant satisfies \(\bar I_1(t,s)\geq3\), its mean obeys \(\bar Z\geq3\), so \(a\geq0\). Furthermore, \(H\geq0\) by construction.
Therefore, the corresponding dissipation relations for the two forms considered are given as follows:
\begin{equation}
	\mathcal D^{(1)}
	=
	\rho_0Nk\,a\,\psi_c'(\bar Z)
	\geq 0
	\qquad
	\text{if }\psi_c'(\bar Z)\geq 0.
\end{equation}

\begin{equation}
	\label{eq:second_moment_dissipation}
	\mathcal D^{(2)}
	=
	\rho_0Nk
	\left\{
	a\left[
	\psi_c'(\bar Z)-\frac{a}{2}\psi_c''(\bar Z)
	\right]
	+
	\frac{H}{2}
	\left[
	\psi_c''(\bar Z)+a\psi_c'''(\bar Z)
	\right]
	\right\}.
\end{equation}
The first-moment approximation is therefore dissipative for any
nondecreasing chain energy. For the second moment approximation, 
dissipation depends on the chosen chain energy.
For the five-term Arruda--Boyce energy in
Equation~\eqref{eqn:AB_5term},
$\psi_c''$ and $\psi_c'''$ are nonnegative for $\bar Z\geq 3$.
The variance contribution is therefore nonnegative.
To bound the remaining contribution, define
$\chi=\bar Z/(3N_c)$. For $0\leq\chi\leq 1$, direct
substitution gives
\begin{equation}
	\begin{aligned}
		\frac{\psi_c'(\bar Z)-\tfrac{a}{2}\psi_c''(\bar Z)}{k_BT}
		&\geq
		\frac{\psi_c'(\bar Z)-\tfrac{\bar Z}{2}\psi_c''(\bar Z)}{k_BT}
		\\
		&=
		\frac12+\frac{3\chi}{20}
		-\frac{513\chi^3}{3500}
		-\frac{42039\chi^4}{134750}
		\\
		&\geq \frac{12899}{67375}>0 \, .
	\end{aligned}
\end{equation}
The final bound follows from concavity of the polynomial on
$0\leq\chi\leq1$, with its minimum attained at $\chi=1$.
Therefore, $3\leq\bar Z\leq 3N_c$ is a sufficient condition
for $\mathcal D^{(2)}\geq 0$ for every nonnegative variance.
States with
$\bar Z>3N_c$ may still have nonnegative dissipation, as determined
directly from Equation~\eqref{eq:second_moment_dissipation}.

For the three-term Arruda--Boyce energy used in the benchmarks,
the same mean contribution satisfies
\begin{equation}
	\frac{\psi_c'(\bar Z)-\tfrac{a}{2}\psi_c''(\bar Z)}{k_BT}
	=
	\frac12+\frac{\bar Z+3}{20N_c}
	+\frac{33\bar Z}{350N_c^2}>0.
\end{equation}
Since $\psi_c''$ and $\psi_c'''$ are also nonnegative, its
second-moment approximation is dissipative without an additional
extensibility bound.


\section{Extension to multiple relative invariants}
\label{app:multiple_invariants}

While the formulation presented in the main text depends on the isochoric relative first invariant, the moment expansion extends to energies involving multiple relative invariants. For generality, we consider the invariants of the full relative right Cauchy–Green tensor, which in three dimensions are given by
\begin{align}
	I_1(t,s)
	&= \mathbf{C}(t):\mathbf{C}^{-1}(s), \\
	I_2(t,s)
	&= \frac{J(t)^2}{J(s)^2}
	\mathbf{C}^{-1}(t):\mathbf{C}(s), \\
	I_3(t,s)
	&= \frac{J(t)^2}{J(s)^2}.
\end{align}
The expression for $I_2$ follows from the identity $I_2=I_3\operatorname{tr}(\mathbf C^{-1})$, applied to the relative right Cauchy–Green tensor. 
The mean invariants are defined by
\begin{equation}
	Z_a(t)
	=
	\int_{-\infty}^{t}\gamma(t,s)I_a(t,s)\,ds,
	\qquad a=1,2,3.
\end{equation}
Their separated representations are
\begin{align}
	Z_1(t)
	&=
	\mathbf{C}(t):
	\int_{-\infty}^{t}
	\gamma(t,s)\mathbf{C}^{-1}(s)\,ds, \\
	Z_2(t)
	&=
	J(t)^2\mathbf{C}^{-1}(t):
	\int_{-\infty}^{t}
	\gamma(t,s)J(s)^{-2}\mathbf{C}(s)\,ds, \\
	Z_3(t)
	&=
	J(t)^2
	\int_{-\infty}^{t}
	\gamma(t,s)J(s)^{-2}\,ds.
\end{align}
Thus, the mean of $I_2$, like that of $I_1$, requires only a second-rank history tensor, while the mean of $I_3$ requires a scalar history variable.

For a coupled chain energy, introduce the invariant vector
$\boldsymbol{\mathcal{I}}=(I_1,I_2,I_3)^{\mathsf{T}}$
and its mean
$\mathbf{Z}=(Z_1,Z_2,Z_3)^{\mathsf{T}}$.
Expansion about $\mathbf{Z}$ gives
\begin{equation}
	W_{\mathrm{TN}}
	\approx
	\rho_0N
	\left[
	\psi_c(\mathbf{Z})
	+
	\frac{1}{2}
	\nabla_{\boldsymbol{\mathcal{I}}}^{2}\psi_c(\mathbf{Z})
	:
	\mathbf{H}
	\right],
\end{equation}
where $\nabla_{\boldsymbol{\mathcal{I}}}^{2}\psi_c$ is the Hessian of the chain energy with respect to its invariant arguments, and
\begin{equation}
	\mathbf{H}(t)
	=
	\int_{-\infty}^{t}
	\gamma(t,s)
	\left[\boldsymbol{\mathcal{I}}(t,s)-\mathbf{Z}(t)\right]
	\otimes
	\left[\boldsymbol{\mathcal{I}}(t,s)-\mathbf{Z}(t)\right]
	\,ds.
\end{equation}
Here, $\mathbf H$ is the covariance matrix of the relative invariants, whose diagonal entries represent the variance of each invariant, while the off-diagonal entries represent cross-covariances that contribute through mixed derivatives of the chain energy.


\subsection{Extension to anisotropic invariants}
The transient network framework can also be extended to  anisotropic structural invariant energies. 
Consider a unit material direction $ \mathbf a_0 $ in the
original reference configuration, and define
\begin{equation}
	I_4(t)
	= \mathbf a_0 \cdot \mathbf C(t)\mathbf a_0,
	\qquad
	I_5(t)
	= \mathbf a_0 \cdot \mathbf C(t)^2\mathbf a_0.
\end{equation}
We assume that a network formed at time $ s $ inherits the normalized,
deformed material direction,
\begin{equation}
	\mathbf a_s
	= \frac{\mathbf F(s)\mathbf a_0}
	{\|\mathbf F(s)\mathbf a_0\|}
	= \frac{\mathbf F(s)\mathbf a_0}{\sqrt{I_4(s)}}.
	\label{eq:structural_formation_direction}
\end{equation}
The preferred orientation is inherited from the deformed material,
while the directional stretch is measured relative to the formation
configuration. Using 
$\mathbf C(t,s)= \mathbf F(s)^{-T}\mathbf C(t)\mathbf F(s)^{-1}$,
the relative structural invariants become
\begin{align}
	I_4(t,s)
	&= \mathbf a_s \cdot \mathbf C(t,s)\mathbf a_s
	= \frac{I_4(t)}{I_4(s)},
	\label{eq:relative_I4}
	\\
	I_5(t,s)
	&= \mathbf a_s \cdot \mathbf C(t,s)^2\mathbf a_s
	= \frac{
		\mathbf a_0 \cdot
		\mathbf C(t)\mathbf C(s)^{-1}\mathbf C(t)\mathbf a_0
	}{I_4(s)}.
	\label{eq:relative_I5}
\end{align}
The separated representations in
Equations~\eqref{eq:relative_I4} and~\eqref{eq:relative_I5}
lead to the history variables
\begin{align}
	b_4(t)
	&= \int_{-\infty}^{t}
	\gamma(t,s)\frac{1}{I_4(s)}\,ds,
	\label{eq:structural_b4}
	\\
	\mathbf A_5(t)
	&= \int_{-\infty}^{t}
	\gamma(t,s)\frac{\mathbf C(s)^{-1}}{I_4(s)}\,ds.
	\label{eq:structural_A5}
\end{align}
The corresponding mean invariants are
\begin{align}
	Z_4(t)
	&= \int_{-\infty}^{t}\gamma(t,s)I_4(t,s)\,ds
	= I_4(t)b_4(t),
	\label{eq:structural_Z4}
	\\
	Z_5(t)
	&= \int_{-\infty}^{t}\gamma(t,s)I_5(t,s)\,ds
	= [\mathbf C(t)\mathbf a_0]\cdot
	\mathbf A_5(t)[\mathbf C(t)\mathbf a_0].
	\label{eq:structural_Z5}
\end{align}
As with the isotropic invariants, these history quantities admit recursive updates under first-order renewal kinetics, allowing anisotropic contributions to be incorporated without explicitly retaining the formation history.

\section{Residual cure and the evolving stress-free state}
\label{sec:Cure}
We consider a simplified curing problem with prescribed kinetics
to illustrate how newly formed chains alter the unloaded
configuration. A full thermodynamic treatment of curing,
including the chemical free energy associated with network
formation, is beyond the scope of this example.
Let $c$ denote the degree of cure and assume
$N(t)=N_\infty c(t)$, with first-order kinetics
\begin{equation}
	\label{eq:residual_cure_law}
	\dot c=k_c(T)(1-c),
	\qquad
	k_c(T)=A_c\exp\left(-\frac{E_c}{RT}\right).
\end{equation}
For a prescribed temperature history, the exact incremental
solution is
\begin{equation}
	\label{eq:residual_cure_update}
	c(t_{N+1})
	=
	1-[1-c(t_N)]
	\exp\left[
	-\int_{t_N}^{t_{N+1}}k_c(T(t))\,dt
	\right].
\end{equation}
Substituting $N(t)=N_\infty c(t)$ into the population weight
defined in ~\ref{sec:Recurrence} gives
\begin{equation}
	\label{eq:cure_population_weight}
	r_N
	=
	\frac{c(t_N)}{c(t_{N+1})}
	\exp(-\tilde k\Delta t),
\end{equation}
where $\tilde k$ is the detachment rate within the increment.
The history tensors are advanced using the same moment
recurrences, while the evolving population $N(t)$ also enters
the prefactor of the network energy and stress.

To isolate chain generation from bond exchange, we consider
a single population with zero detachment rate, $k=0$.
An initially stress-free specimen with degree of cure $c_0=0.4$
is stretched isochorically to $\lambda_c$ over one second at
$250\,\mathrm{K}$. The low ramp temperature suppresses cure
during loading. The stretch is then held for ten seconds at
either $330\,\mathrm{K}$ or $360\,\mathrm{K}$.
For each temperature, $\lambda_c$ is varied from $1$ to $3$
in increments of $0.25$. The illustrative cure parameters are
\begin{equation}
	A_c=10^8\,\mathrm{s}^{-1},
	\qquad
	E_c=60\,\mathrm{kJ\,mol}^{-1}.
\end{equation}
These parameters give final conversions of approximately
$0.564$ at $330\,\mathrm{K}$ and $0.916$ at $360\,\mathrm{K}$,
independent of the imposed stretch.

\begin{figure}[htbp]
	\centering
	\includegraphics[width=0.7\linewidth]{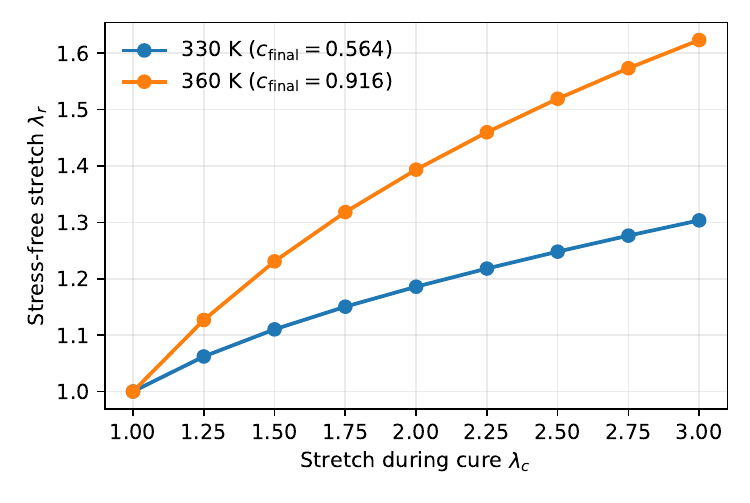}
	\caption{Stress-free equilibrium stretch $\lambda_r$ as a
		function of the imposed curing stretch $\lambda_c$, following
		a ten-second constrained hold at $330\,\mathrm{K}$ or
		$360\,\mathrm{K}$.}
	\label{fig:cure_evolution}
\end{figure}

The mechanical response uses the second-moment, three-term Arruda--Boyce model
with $\mu_\infty=1$ (the instantaneous shear modulus at full cure, $c=1$),
$K_{\mathrm{bulk}}=100$, and $N_c=5$.
After each hold, the stress-free equilibrium stretch $\lambda_r$ is determined
with the final degree of cure and history tensors held fixed.
Figure~\ref{fig:cure_evolution} shows how $\lambda_r$ varies with curing
temperature and imposed stretch $\lambda_c$.
Higher temperatures promote chain generation, while $\lambda_c$ determines
the stress-free configuration of newly formed chains.
The resulting equilibrium stretch reflects the competing contributions
of these chains and those formed in the original configuration.

\bibliographystyle{unsrt}
\bibliography{RTN_a_initials}

\end{document}